\documentclass[twoside]{ilcws07}
\usepackage[latin1]{inputenc}
\usepackage[dvips]{graphicx,epsfig,color}
\usepackage{wrapfig,rotating}
\usepackage{amssymb,amsmath,array} 
\def\eeZh{${\rm e^+e^-}\to{\rm ZH}$}
\def\eeZZll{${\rm e^+e^-}\to{\rm ZZ}\to{\rm \ell^+\ell^-X}$}
\def\Zee{${\rm Z\to\,e^+e^-}$}
\def\Zmm{${\rm Z\to\,\mu^+\mu^-}$} 
\def\eeffff{${\rm e^+e^-}\to4{\rm f}$}
\def\eeff{${\rm e^+e^-}\to2{\rm f}$}
\def\eeZhe{${\rm e^+e^-}\to{\rm Zh}\to{\rm e^+e^-X}$}
\def\eeZhm{${\rm e^+e^-}\to{\rm Zh}\to{\rm \mu^+\mu^-X}$}

\begin{document}
\title{
Prospects to Measure the Higgs Boson Mass and Cross Section 
in \eeZh\ Using the Recoil Mass Spectrum}
\author{W.~Lohmann$^1$, M.~Ohlerich$^{1,2}$, A.~Raspereza$^3$, and A.~Sch\"alicke$^1$
\vspace{.3cm}\\
1- DESY\\
Platanenallee 6, 15738 Zeuthen - Germany
\vspace{.1cm}\\
2- BTU Cottbus \\
Konrad-Wachsmann-Allee 1, 03046 Cottbus - Germany
\vspace{.1cm}\\
3- Max-Planck-Institut f\"ur Physik \\
Foehringer Ring 6, 80805 Muenchen - Germany}

\maketitle

\begin{abstract}
The process \eeZh\ allows to measure the Higgs boson in the recoil mass spectrum 
against the Z boson without any assumptions on the Higgs boson decay. 
We performed a full simulation and reconstruction of \eeZh\ using the 
{\sc Mokka} and {\sc Marlin} packages describing the LDC detector. 
The Z is reconstructed from its decays into electrons and muons. The mass of the Higgs boson is set to $120\,{\rm GeV}$.
Assuming a centre-of-mass energy of $250\,{\rm GeV}$ and an integrated 
luminosity of $50\,{\rm fb}^{-1}$ the Higgs boson mass 
and the Higgs-strahlung cross section can be measured with a precision of $120\,{\rm MeV}$ and 9\%, respectively.
\end{abstract}

\section{Introduction}
The Standard Model (SM)~\cite{sm} predicts one Higgs 
boson as a remnant from the spontaneous symmetry 
breaking mechanism~\cite{higgs}. This mechanism allows 
fermions and the W and Z bosons to acquire their masses 
by interaction with the Higgs field. To discover the 
Higgs boson, therefore, is of crucial interest to complete the SM. 
Electroweak precision measurements suggest the mass of the Higgs 
boson to be of the order ${\cal O}(100\,{\rm GeV})$. Direct searches at LEP 
have set a lower mass limit of $114~{\rm GeV}$. A Higgs boson 
with a mass above $114~{\rm GeV}$
will be accessible in the experiments at the Large Hadron Collider (LHC). 
However, to be sure about the nature of the particle found, it is 
necessary to measure its properties such as mass, width, charge, spin, parity, 
couplings to other particles, and self-couplings to test the internal consistency of the
SM, or to find hints for new physics. 

For the determination of at 
least some of these quantities, LHC will not be sufficient. At the future 
International Linear Collider (ILC), we will have the chance to investigate 
the properties of new particles with high precision in all details. 
This $e^+e^-$ collider with a center-of-mass energy up to $1\,{\rm TeV}$ 
provides a well known initial state, a very clear signature for events of \eeZh, 
and due to its high luminosity sufficient statistics for precision measurements.

In the Higgs-strahlung process \eeZh, we can investigate the coupling of the 
Higgs boson to the Z boson and determine the Higgs boson mass with high 
precision in a relatively model-independent way using the recoil 
mass spectrum against the Z,
\begin{equation}\label{form:recoil}
m_{\rm recoil}^2=s+m_{\rm di-lepton}^2-2\cdot E_{\rm di-lepton} \cdot \sqrt{s}\;,
\end{equation}
where $s$ is the square of the centre-of-mass energy, and $m_{\rm di-lepton}$ and $E_{\rm di-lepton}$ are the mass and the energy of the leptons originating from the Z decay.
 
Previous studies using simplified parametric detector simulations have shown the potential of this technique~\cite{GaLo96}.

Here, we study the prospects to measure the Higgs boson mass and cross section assuming its mass to be $120\,{\rm GeV}$. At a centre-of-mass energy of $250\,{\rm GeV}$, a full detector simulation of the process \eeZh\ is performed using the {\sc Mokka} software package, which simulates events in the LDC detector. The response from the sub-detectors is digitised as in a real experiment and is processed using the {\sc MarlinReco}~\cite{ILCSoft} reconstruction software. In addition, SM background processes are treated in the same way.

The Z is reconstructed from its decays into electrons and muons. Algorithms are developed to 
identify electrons and muons using the tracker and calorimeter information. A likelihood 
technique is used to separate signal events from the SM background processes with high efficiency.

\section{Experimental Conditions}
The study assumes a linear $e^+e^-$ collider operating at a  centre-of-mass energy of $250\,{\rm GeV}$. This energy is chosen because the Higgs-strahlung cross section for the SM Higgs boson with a mass of $120\,{\rm GeV}$ reaches its maximal value.

The statistics for signal and background events corresponds to a luminosity of $50\,{\rm fb}^{-1}$. Signal events were generated using {\sc Pyhtia}~6.4.11 \cite{Sjo06}. Initial and final state bremsstrahlung and beamstrahlung are taken into account. To simulate beamstrahlung the {\sc GuineaPig}~1.4.1 \cite{Sch96} program is used assuming ILC nominal beam parameters. Background events are produced using in addition the event generators {\sc BHwide}~1.04 \cite{Jad97} and {\sc Sherpa}~1.0.10 \cite{GHK04} as listed in Table \ref{tab:xsec}. 

\begin{figure}[!h]
\centerline{\includegraphics[width=0.95\columnwidth]{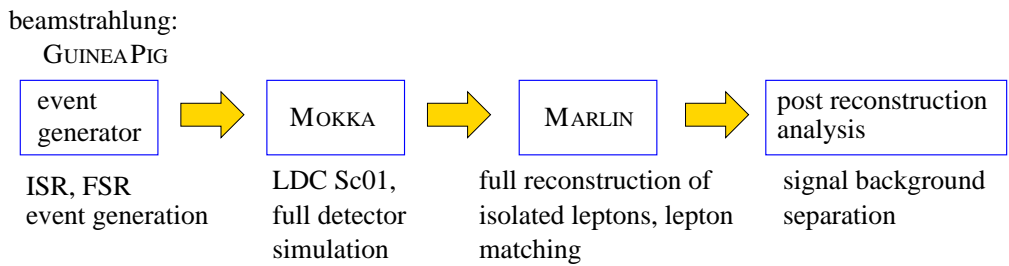}}
\caption{\small Scheme of simulation and analysis stages. 
Events of the different processes are generated using 
{\sc Pythia}~6.4.11 , 
{\sc BHwide}~1.04  and
{\sc Sherpa}~1.0.10.
Beamstrahlung is treated by  {\sc GuineaPig}.
{\sc Mokka} and {\sc Marlin} simulate and reconstruct
events in the LDC detector.
The analysis software is based on ROOT.
}
\label{Fig:fig01}
\end{figure}

Signal and background events are passed through a full detector simulation ({\sc Mokka}) and are processed in the full reconstruction scheme of {\sc Marlin}. A sketch of the simulation stages is shown in Figure \ref{Fig:fig01}.
The LDC detector~\cite{DOD} is used for the simulation and reconstruction. The vertex detector (VTX) consists of five layers of silicon pixel detectors. The main tracker is a TPC of about $3~{\rm m}$ diameter and $4~{\rm m}$ length supplemented by cylindrical silicon strip detectors (SIT) and forward strip and pixel detectors (FTD). The TPC is surrounded by the electromagnetic (ECAL) and hadronic (HCAL) calorimeter, which in turn are enclosed by the $4\,{\rm T}$ magnet and the iron yoke. ECAL is a finely segmented silicon-tungsten sandwich calorimeter. HCAL consists of a steel absorber structure with small scintillator tiles read out with silicon photomultipliers. When this analysis was performed no muon chamber was implemented in {\sc Mokka}. Thus the separation of muons and pions in the particle identification is done using the trackers and calorimeters only. The precision of the momentum measurement of the tracker system (TPC+VTX+SIT+FTD) is obtained to be $\sigma_{p_t}/p_t=7\cdot10^{-5}\cdot p_t\,[{\rm GeV}]$ using the FullLDCTracking processor in {\sc MarlinReco}.
\begin{table}[!h]
\centerline{\begin{tabular}{|rl|r|r|r|}
\hline
&Process & $\sigma\;[{\rm fb}]$ & $N(50\;{\rm fb}^{-1})$ & Generator\\
\hline
1.&$e^+e^-\rightarrow HZ\rightarrow X\ell^+\ell^-$  & $15.0$ & $751$ & {\sc Pythia} \\
2.&$e^+e^-\rightarrow e^+e^-$  & $4144.5$ & $207223$ & {\sc BHwide} \\
3.&$e^+e^-\rightarrow \mu^+\mu^-$  & $4281.0$ & $214050$ & {\sc Pythia} \\
4.&$e^+e^-\rightarrow \tau^+\tau^-$  & $4182.0$ & $209100$ & {\sc Pythia} \\
5.&$e^+e^-\rightarrow W^+W^-\rightarrow Xe,X\mu,Xe\mu$  & $5650.0$ & $282277$ & {\sc Pythia} \\
6.&$e^+e^-\rightarrow e^+e^-f\bar{f}$  & $475.7$ & $23784$ & {\sc Sherpa} \\
7.&$e^+e^-\rightarrow \mu^+\mu^-f\bar{f}$  & $359.4$ & $17970$ & {\sc Sherpa} \\
8.&$e^+e^-\rightarrow e^+e^-e^+e^-$  & $24.6$ & $1231$ & {\sc Sherpa} \\
9.&$e^+e^-\rightarrow \mu^+\mu^-\mu^+\mu^-$  & $7.2$ & $360$ & {\sc Sherpa} \\
10.&$e^+e^-\rightarrow e^+e^-\mu^+\mu^-$  & $177.0$ & $8850$ & {\sc Sherpa} \\
\hline
\end{tabular}}
\caption{\small
The processes simulated for this study, their cross sections, the expected statistics for
an integrated luminosity of ${50\;{\rm fb}^{-1}}$, and the generators used. 
$\ell$ represents $e,\,\mu$ and $f$ stands for $\tau,\nu,q$.
}
\label{tab:xsec}
\end{table}

\section{Simulation Details and Analysis}
\subsubsection*{Signal and Background Processes}
The signal and background processes considered in the analysis, their cross sections and expected statistics at $50\,{\rm fb}^{-1}$, and the programs used to generate events are listed in Table~\ref{tab:xsec}. For the generation of $e^+e^-$ events using {\sc BHwide}, the following cuts are applied: the polar angle range is restricted to $|\cos\vartheta_{\rm lep}|\!<\!0.985$, the electron energy is required to be $E_e\!>\!10\;{\rm GeV}$, the difference of the di-lepton mass and the Z mass is $|m_{ee}-m_Z|\!<\!40\,{\rm GeV}$, and the recoil mass against the Z is in the range $90\,{\rm GeV}\!\le\!m_{\rm recoil}\!\le\!190\,{\rm GeV}$. For event samples generated with {\sc Sherpa}, cuts on the lepton polar angle, $|\cos\vartheta_{\rm lep}|\!<\!0.985$, the di-lepton and di-parton mass, $m_{ee,qq}\!>\!10\,{\rm GeV}$, and the energy of the final state fermions, $E_{\rm fermion}\!>\!5\,{\rm GeV}$, are applied. These cuts avoid divergences in the cross sections, reduce computing time, and would have no influence on the results.

\subsubsection*{Analysis Strategy}

The signature of events from \eeZh\ are two leptons of the same kind and opposite charge. The invariant mass of the two leptons must be in the vicinity of the ${\rm Z}$ mass. The dominant background is expected from the process \eeZZll, which is simulated within the four-fermion processes 6 - 10 in Table \ref{tab:xsec}. Discriminating power is expected from the polar angle distribution of the ${\rm Z}$. Processes 2 and 3 with two electrons or muons in the final state may be selected since initial state radiation leads to a radiative return and thus to an invariant mass near the Z. To distinguish them from the signal, the acoplanarity angle can be used. The process 5 is a possible background due to its high cross section. The polar angle of the leptons can be used to distinguish the signal from this background.

In the fist step of the analysis, we look for isolated electrons and muons in each event using a likelihood method. An electron is identified as an electromagnetic shower in the ECAL that matches the position predicted from a track as impact point into the ECAL. A muon is identified as a track matching to deposits in the ECAL and HCAL compatible with the expectations from a minimum ionising particle.

For the second step, only events with a lepton pair of the same kind and with opposite charge are accepted. If there are several pairings possible, the one with an invariant mass closest to $m_{\rm{Z}}$ is chosen.

For further reduction of the background the following cuts are applied to:
\begin{itemize}
\setlength{\itemsep}{-2pt}
\item[-] the polar angles of the leptons: $|\cos\vartheta_{\rm lep}|\!<\!0.95$,
\item[-] the difference between the invariant di-lepton mass and $m_{\rm{Z}}$: $|m_{\rm di-lepton}-m_{\rm{Z}}|\!<\!30\,{\rm GeV}$,
\item[-] the lepton energy: $E_{\rm lep}\!>\!15\;{\rm GeV}$,
\item[-] the recoil mass:  $90\,{\rm GeV}\!\le\! m_{\rm recoil}\!\le\!190\,{\rm GeV}$,
\item[-] the polar angle of the di-electron system:  $|\cos\vartheta_{\rm di-electron}|<0.90$.
\end{itemize}
The remaining events are analysed using likelihood density functions for the signal, the \eeffff, and the \eeff\ background channels in the variables: acoplanarity angle of the two leptons, acolinearity angle of the two leptons, the di-lepton mass, the polar angle of the di-lepton system, the polar angles of the two leptons and the transverse momentum of the Z. 

For each event a likelihood is calculated characterising its compatibility with a signal event. A cut on this likelihood is set such that the quantity $\frac{\sqrt{S+B}}{S}$ is minimised in the mass range from $119\,{\rm GeV}$ to $125\,{\rm GeV}$. Here, S and B are the numbers of signal and background events, respectively, in the final sample.

\section{Results}
\begin{figure}[!ht]
\centerline{
\includegraphics[width=0.5\columnwidth]{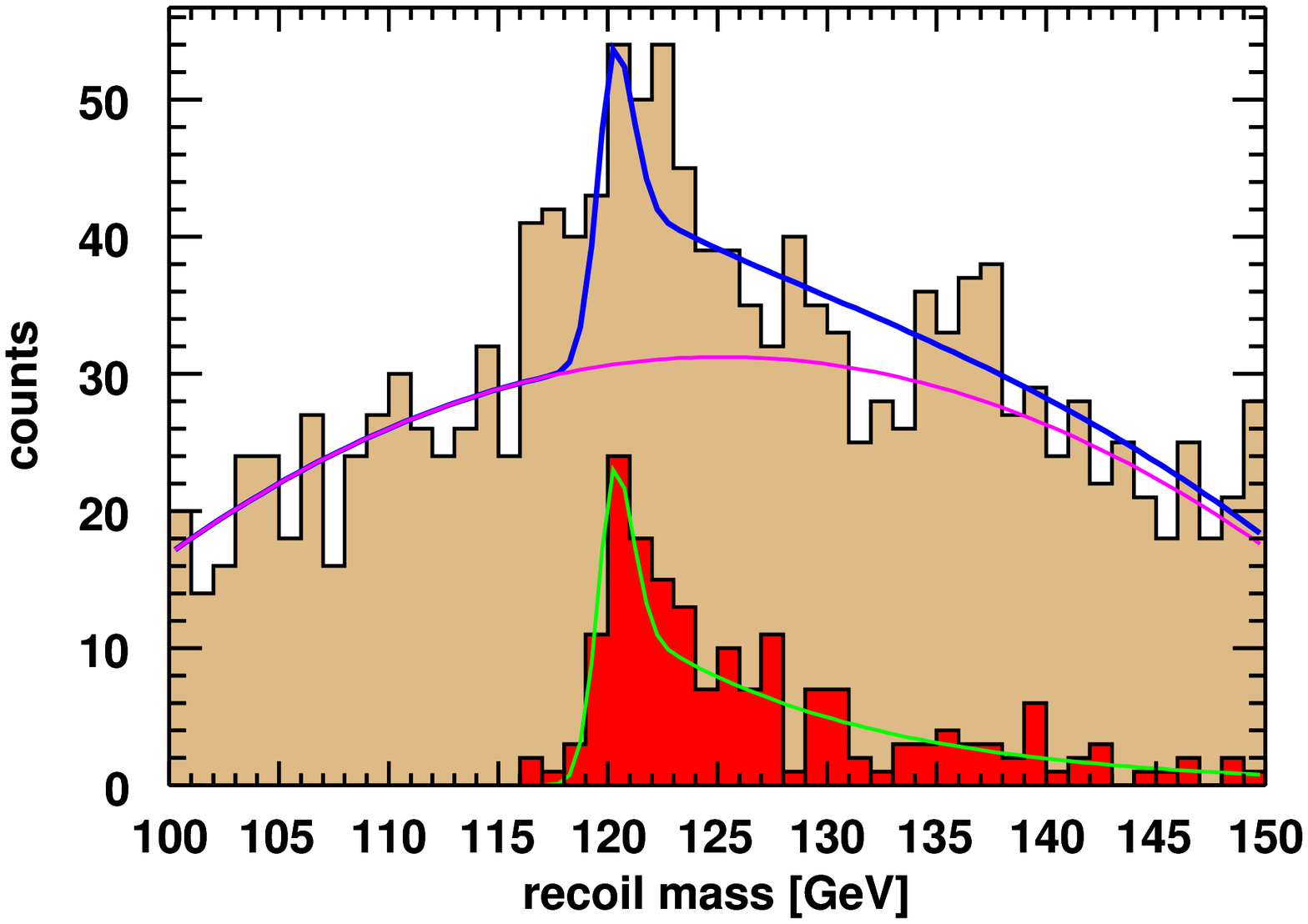}
\includegraphics[width=0.5\columnwidth]{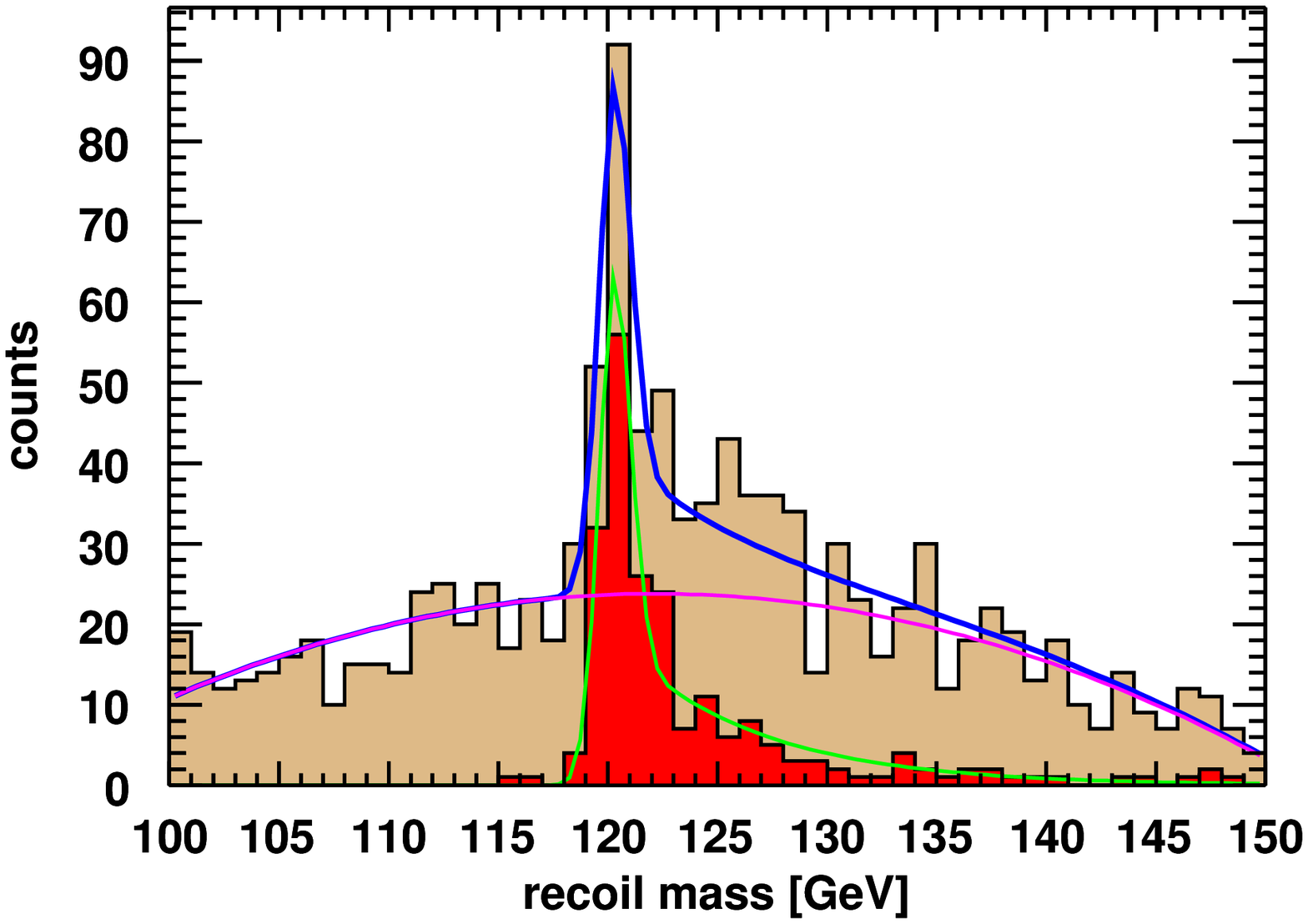}}
\caption{\small 
The recoil mass distributions of the selected \eeZh\ events,
left for \Zee\ and right for \Zmm\ final states, respectively.
The dark red distribution originates from the signal process and the light
red distribution from the remaining background.
}\label{Fig:fig02}
\end{figure}

In the final sample, the signal selection efficiency is obtained to be 43.1\% for the \eeZhe\ and 57.2\% for the \eeZhm\ channel, respectively. The recoil mass spectra are shown in Figure \ref{Fig:fig02} for these processes. In both spectra a signal peak is seen on top of a moderate background. The signal has a smaller width in the di-muon channel, presumably because the muon track measurement is more precise due to less bremsstrahlung in the material of the detectors. This is confirmed by the resolution function for the transverse momentum, which has a pronounced tail to lower reconstructed momenta for electrons.  

\subsubsection*{Cross Section Measurement}
The recoil mass spectra in Figure \ref{Fig:fig02} are used to determine the cross sections for the processes \eeZhe\ and \eeZhm. The background originating from known SM processes is parametrised by a polynomial and kept constant in the fit to determine the amount of the signal. The signal is described using the following parametrisation,
\begin{equation}\label{form:signal}
s(x)={\rm Norm}_{\rm GausExp}\begin{cases}
  e^{-(x-m_{\rm 0})^2/(2\sigma^2_{\rm gaus})} &:\,x\!<\!m_{\rm 0}\;,\\
\beta e^{-(x-m_{\rm 0})^2/(2\sigma^2_{\rm gaus})}+(1-\beta)e^{-(x-m_{\rm 0})/\lambda}&:\,x\!>\!m_{\rm 0}\;,
\end{cases}
\end{equation}
where $m_{\rm 0}$ is the central value of the peak, $\lambda$ a constant to describe the tail to larger values in the signal mass distribution and $1-\beta$ is the fraction of the tail. The tail to larger mass values in the signal is caused by bremsstrahlung and beamstrahlung. The former is well predicted from QED and the latter depends on the machine parameters. For known and reasonable stable machine parameters the shape and fraction of the tail can be determined, and we keep it constant in the fit, varying only $m_{\rm 0}$ and the number of events in the signal, $N_{\rm signal}$. The cross section of the signal process is obtained from
\begin{equation}\label{form:xsec}
\sigma({\rm process})=N_{\rm signal}/({\cal L}\varepsilon),
\end{equation}
where ${\cal L}$ is the integrated luminosity, and $\varepsilon$ 
is the signal selection efficiency. The results obtained are $\sigma(${\eeZh $)=216.0\,{\rm fb}$} 
with an uncertainty of 20\% using the di-electron final state and $\sigma(${\eeZh $)=219.7\,{\rm fb}$}
with an uncertainty of 10\% using the di-muon final state. Both results agree with 
the value of the cross section of $226.8\,{\rm fb}$ obtained from {\sc Pythia}.\\

An alternative method is to count all events, $N$, in the signal mass range from $119\,{\rm GeV}$ to $125\,{\rm GeV}$ and to subtract the background. The latter is obtained from a high statistics Monte Carlo simulation or from the integral over a parametrised background distribution in the same mass range, $\langle B\rangle$. The cross section is then given by
\begin{equation}\label{form:xsec2}
\sigma({\rm process})=(N-\langle B\rangle)/({\cal L}\varepsilon),
\end{equation}
with an uncertainty of $(\pm\;\sqrt{N}/(N-\langle B\rangle)\;[\%]\;$. Using this method no assumption on the signal peak parametrisation is needed. The results obtained are compatible with the fit results given above.

\subsubsection*{Higgs boson mass}

To determine the Higgs boson mass from the spectra shown in Figure \ref{Fig:fig02} a likelihood method is used. Several signal samples of high statistics with Higgs boson masses between $119$ and $121\,{\rm GeV}$ are generated and processed through the full simulation, reconstruction and analysis chain. The obtained spectra are parametrised using Formula (\ref{form:signal}). These parametrisations are then used in an unbinned likelihood fit to the simulated event samples shown in Figure \ref{Fig:fig02} to determine the Higgs boson mass $m_h$. The results are $m_h=119.78\pm0.42\,{\rm GeV}$ and $m_h=120.09\pm0.12\,{\rm GeV}$ for the \Zee\ and \Zmm\ decays, respectively. 
\begin{figure}[!ht]
\centering
\includegraphics[width=0.49\columnwidth]{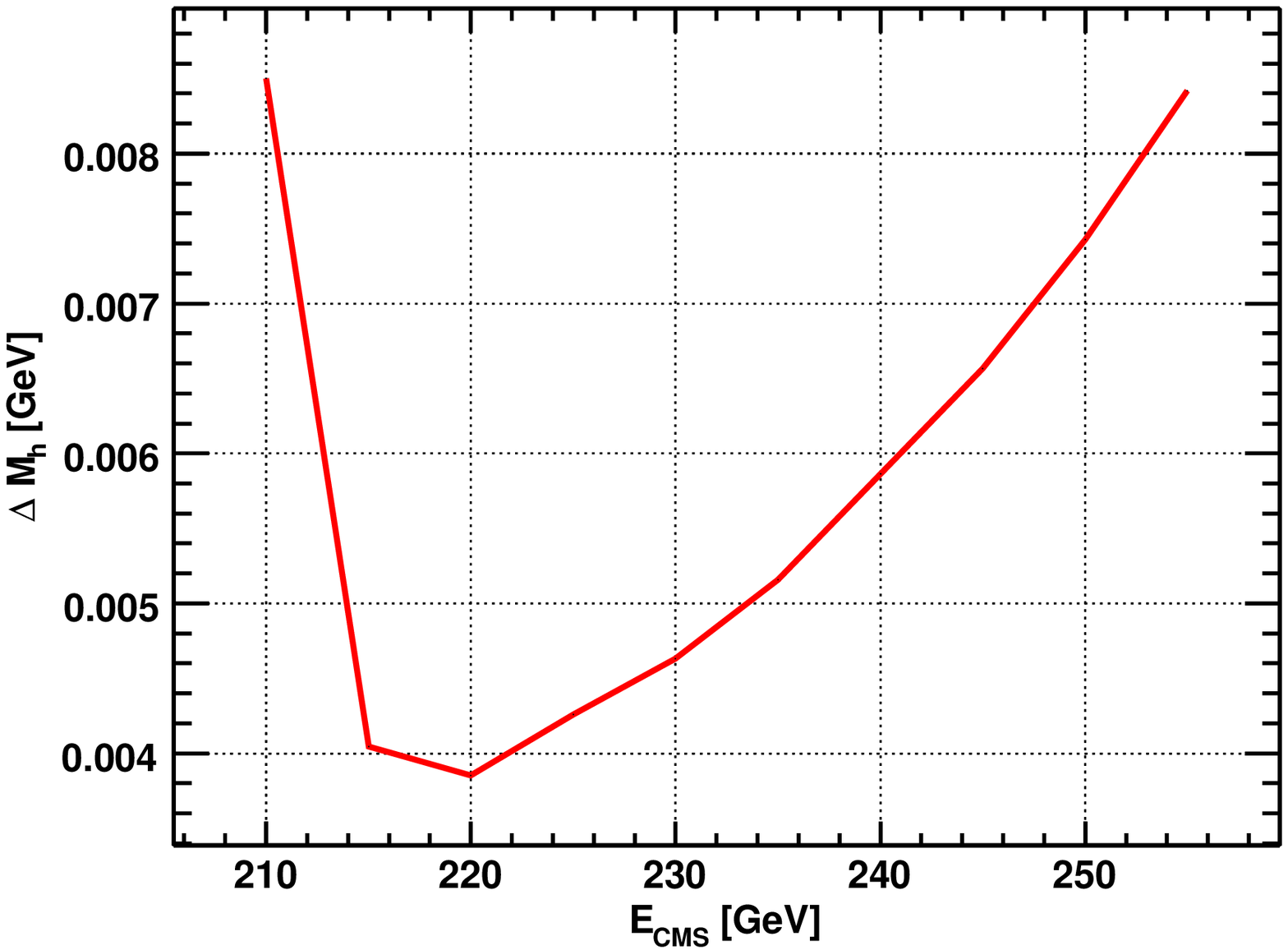}
\includegraphics[width=0.49\columnwidth]{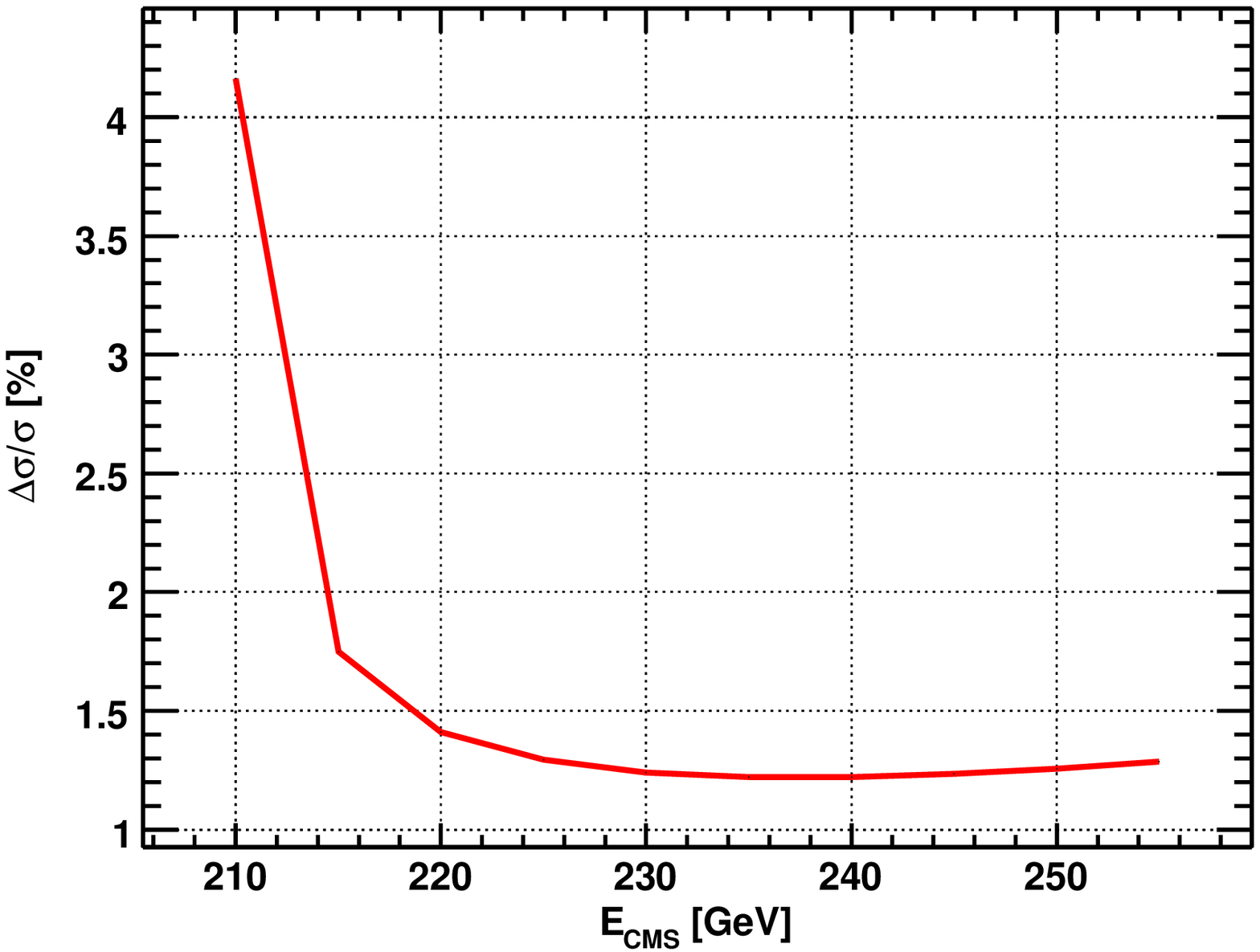}
\caption{\small Uncertainties on the Higgs boson recoil mass measurement (left) and on the Higgs boson production cross section (right) as a function of the centre-of-mass energy. A Monte Carlo toy model with parametrised momentum resolution is used.}\label{Fig:fig06}
\end{figure}

\subsubsection*{Estimate of the Optimal Centre-of-Mass Energy for the Measurements}
A Monte Carlo toy model is used to estimate the optimal centre-of-mass energy 
for the measurement of the Higgs boson mass and cross section. 
The resolution of the track momentum measurement 
is parametrised as $\sigma_{p_t}/p_t=10^{-4}\cdot p_t\,[{\rm GeV}]$. 
A lepton identification and pair matching is performed as described above. 
For a Higgs boson mass of $120\,{\rm GeV}$ the recoil mass spectra are 
used to determine the Higgs boson mass and cross section for
centre-of-mass energies between $210\,{\rm GeV}$ and $250\,{\rm GeV}$.
The estimated accuracies for the mass and cross section measurements are 
shown in Figure \ref{Fig:fig06} as a function of the centre-of-mass energy assuming an integrated luminosity of $500\,{\rm fb}^{-1}$.
The minimal uncertainty on the recoil mass occurs for $E_{\rm cms}=220\,{\rm GeV}$, 
in agreement with the results obtained in Ref.\cite{philip}. 
The cross section uncertainty is minimal for $E_{\rm cms}=240\,{\rm GeV}$, 
becoming worse by about 20\% at  $E_{\rm cms}=220\,{\rm GeV}$.

\section{Conclusions}

The recoil mass technique is a unique tool to determine the mass and cross section of the Higgs boson at the ILC. For the first time the prospects obtained from a full detector simulation and reconstruction are presented. Choosing the center-of-mass energy a few ten GeV above the kinematic threshold, here $250\,{\rm GeV}$ for $m_h=120\,{\rm GeV}$, the Higgs boson mass and cross section can be determined with an accuracy of 120 MeV and 9\%, respectively, using only $50\,{\rm fb}^{-1}$. To reach a similar accuracy at a center-of-mass energy of $350\,{\rm GeV}$, an integrated luminosity $500\,{\rm fb}^{-1}$ is needed.

The talk held at the LCWS is available under Ref.~\cite{url}.

\begin{footnotesize}

\end{footnotesize}


\begin{thebibliography}{99}

\bibitem{sm} S.L. Glashow, Nucl. Phys. {\bf 22} (1961) 579;
             S.~Weinberg, Phys. Rev. Lett. {\bf 19} (1967) 1264;
             A.~Salam, {\em Elementary Particle Theory}, edited by N.~Svartholm
             (Almqvist and Wiksell, Stockholm, 1968), p. 367.

\bibitem{higgs} P.W. Higgs, Phys. Lett. {\bf 12} (1964) 132, Phys. Rev. Lett. {\bf 13} (1964) 508 and Phys. Rev. {\bf 145} (1966) 1156 ;
                F.~Englert and R.~Brout, Phys. Rev. Lett. {\bf 13} (1964) 321;
                G.S.~Guralnik, C.R.~Hagen and T.W.B.~Kibble, Phys. Rev. Lett. {\bf 13} (1964) 585.

\bibitem{GaLo96} P.~Garcia-Abia and W.~Lohmann,
  Eur.\ Phys.\ J.\ direct C {\bf 2} (2000) 2
  [arXiv:hep-ex/9908065].

\bibitem{ILCSoft} ILC Software Portal: \verb$http://www-flc.desy.de/ilcsoft$

\bibitem{Sjo06}
  T.~Sjostrand, S.~Mrenna and P.~Skands,
  JHEP {\bf 0605} (2006) 026
  [arXiv:hep-ph/0603175].

\bibitem{Sch96}
   D.~Schulte, 
   Universit\"at Hamburg (1996)

\bibitem{Jad97}
  S.~Jadach, W.~Placzek and B.~F.~L.~Ward,
  Phys.\ Lett.\  B {\bf 390} (1997) 298
  [arXiv:hep-ph/9608412].

\bibitem{GHK04}
  T.~Gleisberg, S.~H\"oche, F.~Krauss, A.~Sch\"alicke, S.~Schumann and J.~C.~Winter,
  JHEP {\bf 0402} (2004) 056
  [arXiv:hep-ph/0311263].


\bibitem{DOD}
  DOD, LDC Working Group: \verb$http://www.ilcldc.org/documents/dod$

\bibitem{philip}
  F.~Richard and P. Bambade, LAL-Orsay 07/03.

\bibitem{url} Slides: \verb$http://ilcagenda.linearcollider.org/materialDisplay.py?contribId=159&materialId=sli$\\
\verb$       des&confId=1296$
\end{thebibliography}
\end{document}